\documentclass[conference]{IEEEtran}
\usepackage[utf8]{inputenc}
\usepackage[T1]{fontenc}
\usepackage{longtable}
\usepackage{enumitem}
\usepackage{graphicx}
\usepackage{grffile}
\usepackage{array}
\usepackage{longtable}
\usepackage{wrapfig}
\usepackage{rotating}
\usepackage[normalem]{ulem}
\usepackage{amsmath}
\usepackage{textcomp}
\usepackage{amssymb}
\usepackage{capt-of}
\usepackage{hyperref}
\usepackage{quoting}
\usepackage{supertabular}
\usepackage{epigraph}
\author{
\IEEEauthorblockN{Curtis McCord}
\IEEEauthorblockA{University of Toronto\\Faculty of Information}
\and
\IEEEauthorblockN{Christoph Becker}
\IEEEauthorblockA{University of Toronto\\Faculty of Information}
}

\date{\today}
\title{Sidewalk and Toronto: \\ Critical Systems Heuristics and the Smart City}
\begin{document}

\maketitle
\thispagestyle{plain}
\pagestyle{plain}
\begin{abstract}
`Smart cities', urban development projects that design computational systems and sensory technology to monitor activity and regulate energy consumption and resource distribution, are a frontier for the prospective deployment of ICTs for sustainability. Often reduced to technological problems of optimization, these projects have implications far beyond narrow environmental and consumptive frames of sustainability. Studying them requires frameworks that support us in examining technological and environmental sustainability dimensions jointly with social justice perspectives. This paper uses Critical Systems Heuristics (CSH) to examine the design of Sidewalk Toronto, an ongoing smart city development. We explore how the professed values guiding the project are contentiously enacted, and we argue that key stakeholders and beneficiaries in the planning process significantly constrain the emancipatory and transformative potential of the project by marginalizing the role of residents in determining project purposes. This analysis contributes an example that illustrates the relevance of critical systems thinking in ICT4S and offers CSH as a conceptual frame that supports critical reflection on the tensions between the visions and realities of `sustainable' ways of organizing human life. 
\end{abstract}

\section{Introduction}
\begin{quoting}[font=itshape,vskip=.5em]{
\noindent`A neighbourhood from the internet up'~\cite{doctoroff2017vision}\newline
`Google’s Guinea-Pig City'~\cite{sauter_googles_2018}\newline
`Smart City, Dumb Deal'~\cite{editorial_guardian_2018}
}\end{quoting}

Controversy surrounds the proposed development of the Sidewalk Labs `smart city' project in Toronto. In the world's most diverse city, the world's most powerful computing business (Google's Alphabet) has partnered with the municipal government, promising to build the sustainable city of the future. This project is imagined as a prototype, a technology-driven flagship, and a vehicle for legitimizing an ICT company's bid to shape the future of cities worldwide.

The term `smart city' is full of promises: as human activity condenses into urban environments, urban life has become the site for sustainable design. Since ICT have become a foundation of dominating cultures and economies, they are seen as a choice strategy to make cities sustainable. The drivers of this kind of sustainable development are large-scale collection and analysis of observational and statistical data, and cybernetic feedback through embedded devices and human-facing software. The term `smart city' subsumes these technologies, the logic that relates them to goals of sustainability, and the aesthetic of a sustainable city.

This paper examines how the Sidewalk Toronto project's purposes are influenced by the goals of it's most powerful stakeholders, Waterfront Toronto and Sidewalk Labs. This paper will argue that sustainability, and thus smart city projects, should be considered more holistically than is possible through the narrow lenses of technological optimization and environmental sustainability. Failure to consider systems critically can leave out considerations with important impacts for how sustainability is pursued, and how power and decision-making might influence that pursuit.

The paper's argument continues the trajectories set by Mann et al. \cite{mann2018shifting}, Easterbrook \cite{easterbrook2014computational} and Becker et al. \cite{becker2015sustainability}, namely that Information Communication Technology for Sustainability (ICT4S) research must take a more ambitious, critical, and holistic approach to sustainable design than is possible through piecemeal interventions or optimization of environmental parameters \cite{mann2018shifting}. Sustainability design has implications for our economies, societies, technologies, our cooperative work, and our individual lives\cite{becker2015sustainability}. Building on Easterbrook's critique of ``computational thinking''\cite{easterbrook2014computational}, we explore the viability of systems thinking concepts by analysing Sidewalk Toronto, a high profile sustainable smart city proposal. Unlike Easterbrook's focus on System Dynamics, however, our analysis uses Critical Systems Heuristics (CSH) to examine the planning and engagement process of Sidewalk Toronto, and to mount a \textit{boundary critique} that provides insights on the value judgments and justifications that promote and legitimate the project's technology choices and designs. By tracing the concerns of those involved and those affected, we examine how values and interests influence and constrain the purposes and vision of Sidewalk Toronto, offering CSH as a conceptual framework to support technology-supported transitions to just and sustainable societies. We hope this can help the ICT4S community to better understand how each of us can `shift the maturity needle' upwards~\cite{mann2018shifting}. 

\section{Background}
\subsection{Smart Cities in ICT4S}
Smart city research in the ICT4S community has focused on the design of specific products to affect consumption patterns \cite{kamilaris2015social,von2015citizen}, to provide ICT-based management and evaluation tools \cite{shahrokni2014big,brylka2014ict,kramers2018aaas}, as well as to understand the new relationships between technologists and policymakers forged in smart city projects \cite{cosgrave2014smart,kramers2014planning}, even making structural critiques about the agency of citizens herein \cite{borjesson2015ict}.

As the concept of the smart city becomes more popular in the ICT4S community, care must be taken to avoid replicating the weaknesses in considering technological systems as separable from the much larger and complex systems of social organization and reproduction within which they are embedded. For example, Borjesson et al. argue that a narrow technological/environmental frame for smart cities neglect the importance of social-systemic patterns of consumption and activity, in favour of a simplistic understanding of humans as atomized beings who make decisions based on economic and rational calculations \cite{borjesson2015ict}. Kamilaris et al. \cite{kamilaris2015social} and von Heland et al.\cite{von2015citizen} discuss interventions that build upon economic and social relations among their participants, although these relations also act as a source of inertia.

To consider smart cities as abstract systems that can be algorithmically optimized for sustainable energy and resource use fails to do justice both to present and future residents and to the ideal of sustainability. As Mann et al. argue, approaches to sustainability in the ICT4S community must make a strident effort to move beyond merely acknowledging the importance of sustainability, or simply proposing product-based interventions or efficiency finding manoeuvres \cite{mann2018shifting}. For us, this means enabling ICT4S research to conceive of sustainability holistically, without abstracting it to information and resource management projects. Transformative sustainability requires critical analysis and questioning of much of humanity's habits and practises, be they social, economic, political or technological. Nowhere is this more apparent than in the discourse of the smart city, a new construction of urban life according to state of the art technologies and practises. At this intersection, the political, economic and social dimensions of sustainability practise are readily implicated, and researchers committed to sustainability require tools to critically interrogate these relationships.

\subsection{Sidewalk Toronto}
The case we analyze is the Sidewalk Toronto smart city project being planned in the 12-acre Quayside area of Toronto's waterfront: a joint venture between Waterfront Toronto (WT) and Sidewalk Labs (SL), an urban development and technology firm and Alphabet (Google's parent company) subsidiary. Sidewalk Toronto is to be a pilot for future local development, and a test-bed for smart city development globally. In 2017, WT issued a request for proposals that situated smart cities as a technological approach to sustainability, eliciting a partnering firm that could use ICTs to create a ``climate positive approach [to urban design] that will lead the world in city building practises'' \cite{waterfront2017rfp}. This \emph{RFP} was answered and won by an ambitious \emph{Vision} statement by SL for an ecologically sustainable community built on terms of cybernetic ecology that could serve as a replicable and universalizable model for smart city projects globally. SL's \emph{Vision} seeks to sustainability and replicability by building a neighbourhood informated and monitored at all levels, the ``most measurable community in the world'' \cite[p.22]{doctoroff2017vision}.

Excepting the foregrounding of sustainability concerns, the principles extolled in WT's \emph{RFP} and Sidewalk's \emph{Vision} are congruent with the ``Smart City Principles'' explored by Cosgrove et al., placing the focus of human work on service provision driven by optimization, grounded in an ``information marketplace'' \cite{cosgrave2014smart}. Beyond environmental sustainability, Sidewalk Toronto's ecology includes social and economic dimensions of sustainability through its focus on ``complete communities'', and through specific products like a ``public realm management system''. The design process itself strives for ``holistic planning'', where ``innovation, community priorities, policy objectives, placemaking, phasing, infrastructure, economics, market, site planning, and technical issues will be thoughtfully merged'' \cite[p.59]{doctoroff2017vision}. This public engagement component is the front line in drawing system boundaries that will structure the smart city and its operation.

\subsection{Systems Thinking in ICT4S}
Sidewalk Toronto is envisioned in terms of systems: a ``next-generation transit system'', a ``district wide energy system'', an ``ecosystem'' that supports economic agents, as well as a collection of information systems that support relations between residents, such as the ``public realm management system'' \cite{doctoroff2017vision}. The \emph{Vision} even foresees groups of people engaged in everyday life systemically: ``a system of networked neighbourhoods\dots [that] will begin to operate at a system scale, like the internet, generating advantages that increase with each new node'' \cite[pg.21]{doctoroff2017vision}. Systems Thinking is essential for a critique of Sidewalk Toronto because the project pursues aims of sustainability through the systematization of everyday life, creating a space where the activities are monitored as informational transactions, refined into actionable intelligence, and turned back onto the behaviours of smart city residents. As a systems design project, Sidewalk Toronto intends to construct a system of life from the ground up, so that it might be replicated universally.

Easterbrook's call for the integration of systems thinking contexts into computational research stems from three perceived weaknesses in the dominating `computational' way of thinking. 

First, he argues that the domain ontology of computational thinking is problematically biased by its dependence on computational terms, techniques, metaphors, and heuristics for describing the world \cite[p.239-240]{easterbrook2014computational}. Computational thinking is most powerful when complexity can be reduced to deterministic sets of variables and interactions, managed hierarchically by a system and ``solved'' by reckoning. As Easterbrook notes, though subfields in computer science have developed techniques to capture what was overlooked, they are for the most part expansions of computational thinking. To enrich the descriptive capacities of computational thinkers, Easterbrook proposes the use of concepts from the area of \textit{System Dynamics}, which has a close connection of ecological thinking and uses concepts of feedback, stocks, and flows. 

Second, Easterbrook argues that computational thinking has a limited capacity to understand how systemic change occurs. Either change is explained deterministically (in terms of having access to technologies or information previously absent), or as the result of responsible individuals, who ``have agency over their social and environmental impacts, ... [and only need] better tools to help them become more sustainable'' \cite[p.240]{easterbrook2014computational}. In thinking about smart cities, we try to avoid technological solutionism. 

Third, Easterbrook criticizes computational thinking as ill-equipped to handle complexity \emph{critically}, and as a result struggles to consider ``questions about how relationships of power are created and maintained in society, and how the tools that mediate social interactions affect these relationships\dots [and] who has the power to create or prevent change'' \cite[p.241]{easterbrook2014computational}. These questions are essential to considering how our societies can be reorganized to be holistically sustainable. Easterbrook here covers the key presuppositions that underpin a critical systems approach: \emph{(1)} that deciding on an account of a system necessarily means partitioning it from the larger context in which it is embedded, \emph{(2)} that ``any interesting system'' will be complex enough such that no single definitive model or counter-factual claim can be non-probabilistically true (so disagreement is inevitable), and \emph{(3)} that the standpoint of the observers are undeniably mediated by the ways we think, act and learn about a situation \cite[p.242]{easterbrook2014computational}. 

\section{Critical Systems Heuristics}
As Ulrich compellingly argues, it is not enough to try and be holistic. Systems thinkers must also deal critically with their own inevitable selectivity and lack of comprehensiveness, by reflecting on their own understandings as equally partial. 

As the major framework in critical systems thinking\cite{jackson2003systems}, CSH is of course \textit{systemic}. Some kinds of systems thinking --including System Dynamics approaches-- are just as focused as computational thinking on description, abstraction, and modelling. By contrast, CSH is not dependent on a `realist' ontology that assumes descriptions of a system correspond to a real (existing) arrangement in the world. CSH is concerned with \emph{discursive} acts, with decisions made by multiple parties with varying goals, epistemic frameworks, and ways of describing the world and the systems being designed. Rather than seeking to classify the component elements of an assumed system or provide a model of their relations, CSH focuses on the reflexive consideration of a designed system's purpose or goals, and how these are justified by a `reference system' of assumptions and judgments. The central entrance point to this reflection is the system's purpose, which is not a thing in itself, but deployed by someone as a matter of heuristic necessity. Any system or plan humans design will be designed for some purpose, to serve some interest or need, according to some worldview \cite[p.243]{ulrich1983critical}. Purposes are necessary to make an endeavour intelligible.

CSH is \textit{critical}. In contrast with structural accounts of change, critical systems thinking means not only acknowledging the systemic interconnections of behavioural patterns (such as the use of fossil fuels with the design of cities), but to see that these patterns are not homeostatic but \emph{actively maintained} and could, therefore, be changed.
\begin{figure*}[t]
    \centering
    \includegraphics{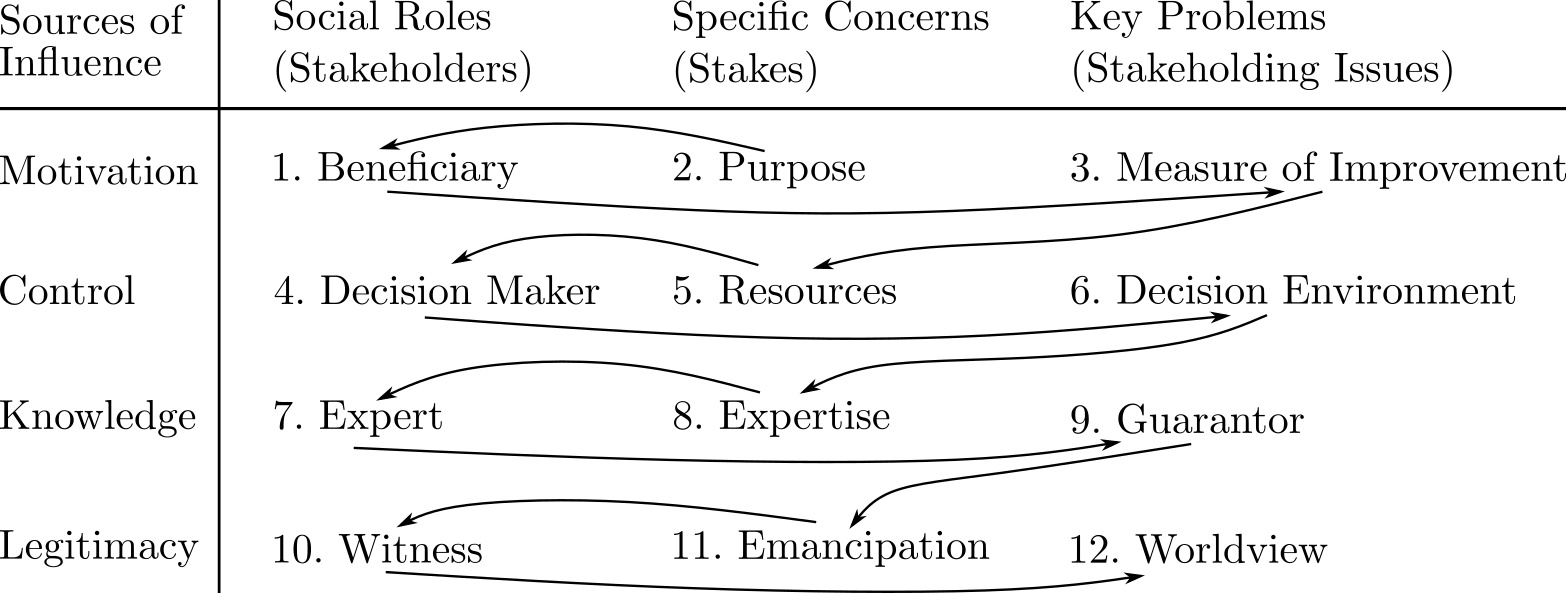}
    \caption{The CSH categories with a suggested order adapted from\cite{ulrich2010critical,reynolds2007evaluation}}
    \label{fig:csh}
    \vspace{-0.3cm}
\end{figure*}

CSH is \emph{heuristic}, admitting that no standpoint or theory can ever sufficiently justify its own assumptions \cite[p.287]{ulrich1983critical}. Any analysis, including our own, must take its own partial standpoint, and can neither comprehensively describe a situation, nor subsume all possible perspectives of it. At best, CSH can seek to reveal and problematize the normative assumptions informing a plan, making clear the contingent nature of these boundaries and making them the subject of deliberation. CSH is intended to enable citizens or participants in a decision making process to engage in critique of expert knowledge, and to discover sources of deception or implicit strategic action \cite[p.22]{ulrich1983critical}.

CSH aims to avoid coercion in planning by, inasmuch as possible, eroding the boundaries between those \emph{affected} by a system and those \emph{involved} in the decision making process \cite[p.248]{ulrich1983critical}. At stake is how those involved, as participants and observers, can rationally justify and legitimate the boundaries and concepts that structure systems design. Domain or technical experts may be better able to make claims or have more power to make decisions, but they cannot justify or legitimate these claims without recourse to assumptions about the way the world is structured and how a system ought to be designed to accomplish the purposes for which it is created. Only the affected can legitimate the implications of any technology design as far as they affect their own lived experience. On this matter, they are the legitimate experts.

Those involved in decision making construct the boundary judgments that constitute an intelligible social systems design. This raises questions about motivation, control, and the expertise needed for implementation. The sorts of decisions inevitably made to plan a system are summarized in twelve boundary judgments derived from the intersection of four categories with three levels of concern \cite[p.244ff]{ulrich1983critical}. These four categories are shown in Fig.\ref{fig:csh}. Application of CSH involves moving through the twelve questions, often shifting between ``is'' and ``ought'' modes. The former mode asks us to reflect on how we see boundary judgments in practise. The latter is intended to stimulate critical reflection on the adequacy of those judgments. It might focus on what is left out, and the moves made by the involved to constrain or preempt boundaries for systems design.  

CSH can be used: (1) to support ideal planning or critical reflection in reflective practise or action research; (2) as an evaluative framework applied to planning situations or decision making processes that define and specify a system to be designed \cite{reynolds2007evaluation}; and (3) polemically, to question experts' claims about what is `objectively necessary' and expose the implicit boundary judgments they make. CSH is not simply a questionnaire to be populated, however, but a system of categories to structure a discursive, reflective, or dialectic process. As such, its raises questions rather than answers them. Our use of CSH to critically examine the boundary judgments in Sidewalk Toronto is no exception. For this analysis, we used official SL and WT documentation, public engagement reports, and press articles. We have also participated in several planning sessions, and have visited the affected spaces. 

\section{CSH and the Smart City}
\subsection{The official story}
The project's stated \textbf{purpose} is to create  ``a vibrant, climate-positive and prosperous community\dots as a national and global model to encourage market transformation towards climate-positive city building'' \cite{waterfront2017rfp}. The \emph{RFP} positions the initial site (Quayside) as stepping stone for ``subsequent developments on the eastern waterfront''\cite[p.6]{waterfront2017rfp}). WT adopts the ``triple bottom line'' approach that balances environmental, economic, and ``socio-cultural'' purposes in the ``3Ps'' of people, profits, and planet \cite[p.A4]{waterfront2017rfp,henriques_triple_2004}.

Declared \textbf{beneficiaries} include prospective \textit{residents} in Toronto, including ``people of all income levels and at all stages of life'' who will benefit from a ``future proof'' life in an inclusive ``complete community''. This includes ``employers and job creators'' who benefit from access to a dense cluster of skilled labour and competing firms, as well as the businesses who sustaining and sustained by the community. \textit{Tourists and residents} of the greater Toronto area will benefit from the addition of significant cultural and recreational amenities to the waterfront\cite{doctoroff2017vision,waterfront2017rfp}. Finally, the \emph{RFP} clearly denotes the prospective \textit{partner} (SL) as a beneficiary to balance the requirement that the partner provide \$50M CAD to cover project costs. Key benefits to SL are an ``unparalleled testbed environment'' to ``showcase advanced technologies, building materials, sustainable practices and innovative business models that demonstrate pragmatic solutions to climate positive urban development''; ``financial opportunities'' from real-estate transactions (even beyond Quayside); and, critically, intellectual property (IP) \cite[p.18]{waterfront2017rfp}.

To assure that the project has succeeded in its purpose to create an environmentally and socially sustainable community in a market friendly urban area, \textbf{measures of improvement} (MoIs) are developed. These provide an important basis to examine how parties involved in decision making understand the system's purposes and beneficiaries. WT's \emph{RFP} sets out sustainability-related MoIs in their \emph{Resilience and Innovation Framework} and asks potential partners to provide key performance indicators to measure and evaluate success \cite{waterfront2017rfp}. A collection of these can be found in the Technical appendix to the \emph{Vision} document \cite[pgs.162-171]{doctoroff2017vision}. Concerning social sustainability, WT sets measures for community creation in terms of ``quality of life'', measured in part through minimum shares for affordable housing (20\% of units), ``convenient and efficient'' transit options, as well as the provision of sufficient social and cultural amenities and high quality design. For WT, the purpose of building ``complete communities'' is realized by meeting people's needs for jobs, services, housing, community infrastructure, as well as transit options. 

Before the release of the major planning document, the \emph{Master Innovation and Development Plan} (MIDP), the \emph{Plan Draft Agreement} (PDA)\cite{waterfront2018plan} sets out the major \textbf{resources} at stake: land and real-estate property, IP, and data. It  makes clear that no transfers of real-estate to Sidewalk have been made: WT continues to assert its role as a steward of public lands and act as negotiating partner (under the auspices of government) in cases where privately owned land will be acquired within the scope of the MIDP  \cite[p.5]{waterfront2018plan}.  The PDA identifies two types of IP: ``planning materials'' and ``products and services'', referring to user-facing software and applications as well as standard and data layer ``enabler'' technologies\cite[p.41]{waterfront2018plan}. Importantly, data may become a source of IP in the future. Within the ambit of ``data governance'', the PDA insists on compliance with existing laws and encourages design principles such as ``personal privacy, civil society protections and technological sovereignty'' and ``data governance and stewardship that ensures both data/information sovereignty protection and innovation''. An expert advisory board oversees these parts of the plan \cite[p.46-9]{waterfront2018plan}. These commitments reference distributions of control through consent, the creation of an independent data trust, and innovation through open architectures. The PDA establishes WT and SL as joint \textbf{decision makers} about the MIDP's structure and its approval, which will be based on a framework developed by WT\cite{waterfront2018plan}. The ultimate sign-off for the plan lies with the municipal and provincial governments, which constitutes the administrative decision making \textbf{environment} over which the decision makers have no control.

The planning process is centered on public engagement and what the /emph{Vision} calls``holistic planning''. The technical expertise of Sidewalk Members and the facilitation skills of WT and contracted firms continuously collect feedback and responses from citizens. Sidewalk Toronto thus involves an array of \textbf{experts and expertise} in planning, implementation and operational phases. As the MIDP is drafted, pieces of the plan are presented for public comment and engagement on specific ideas. In line with Borjensson et al.'s\cite{borjesson2015ict} call to include lived experience within the planning scope of smart cities systems, the engagement mechanisms deployed by WT and SL have leveraged both residents and experts. Resident Advisory Panels provide  situated everyday expertise in the form of reports \cite{residents2018report}. In Expert Advisory panels, independent subject matter experts contribute to planning within areas such as information management and privacy. Additionally, a fellowship program leverages the experiences and training of students and other young subject matter experts to provide a report and recommendations after visiting a number of cities \cite{fellows2018report}. A summer camp even engaged children to provide experiential knowledge \cite{kids2018report}. 

In principle, the opportunities for residents and members of publics to learn about and contribute to planning supports the broad participation of a wide range of experts. Hundreds of people attend the frequent public meetings and open events, staffed by dozens of volunteers. Public Roundtables feature general and breakout sessions which are recorded and streamable\cite{sidewalk2018youtube}, while charettes and design jams address specific themes with public and expert cooperation. To ask about a \textbf{guarantor} for this project in this light is to ask \emph{what assures that our assumptions are justified?} and, crucially, \emph{what assumptions underpin the credibility of experts?}, including assumptions about the relevance of their knowledge, the correctness of their predictions, and the legitimacy of their perspectives. As purposeful human activity, systems design in the view of CSH \emph{implies} recourse to a guarantor: something in which people must put their trust at the place when the chain of justification breaks off. For example, we might trust in WT's mandate to act in the public interest and believe that market-led innovation and development will guarantee the best or most serviceable smart city. Or, we might trust in the expertise of SL, believing that the state-of-the-art technologies offered by one of the world's most well-known firms is the most effective path to sustainable urban design, and that their careful consideration of their public engagement work will assure public have their say.

While categories of Motivation, Control and Knowledge refer to those involved, Legitimacy addresses those affected. The legitimate inclusion of multiple perspectives requires the \textbf{emancipation} of the affected, the ability to speak their concerns freely, and the obligation of the involved to consider them in good faith. This is to be made possible through commitments to public engagement in the planning process. For example, participants at Roundtables represent themselves; they give their own opinions and thoughts, speaking to specific questions in structured and facilitated venues. The resulting material is collected, analyzed and responded to in iterated representations of the emerging plan. However, public engagement processes like this are clearly designed to establish the democratic legitimacy of the project by ensuring that those affected are represented. To what degree does this self-selected participation in highly structured exercises bring forth an honest \textbf{witness} to testify to the experience of those affected by the outcomes of this process?  In order to examine to what degree those affected are free to emancipate themselves \textit{from the project's} \textbf{worldview}, and to offer their own perspectives, we turn to the conceptual framework of boundary critique and explore how CSH supports it.

\subsection{Boundary critique}
Having used CSH to represent the stated `is' situation, we  discuss selected themes that illustrate how CSH helps us to: (1) \textbf{elicit} and make visible the reference system of assumptions that underpin the project; (2) \textbf{contrast} it with complementary perspectives on \textit{how things are} as well as \textit{how things should be}; (3) \textbf{evaluate} how competing interpretations are politically marginalized through the design of the engagement process; (4) \textbf{identify} and critique central boundary judgments; and (5) effectively \textbf{structure} our own critique so that its normative implications become transparent.
\subsubsection{Elicit reference system} 
CSH can provide a discursive grounding for critique of the normative assumptions at play in the planning process. The first step is the reconstruction of the stated boundaries of the system within the CSH framework. We have done this above through reference to, and analysis of, official documentation and presentations. Placing these statements of value and purpose into the relations of CSH allows us to sketch out themes and boundaries based on our own standpoints, such that we can explore how boundaries are constructed, and potentially in disagreement with their motivating principles.

CSH considers two kinds of boundary judgments. The first kind refers to boundaries are established more or less explicitly in the process of partitioning a system, or in deciding what should be considered a component or relation of the system and what is considered as environmental. The second kind refers to the boundary judgments supporting the first. Any decision about what and how to consider a system is a claim that admits of argumentation and justification. These supporting statements are also claims that admit of justification, going on until justification stops with a rhetorical appeal to something like expertise or to a view of reality. The point at which a boundary admits of varying interpretation from complimentary perspectives, or when the reason to accept it is left as an open question, constitutes a ``justification break-off''\cite{ulrich1987critical}. Revealing these justification break-offs is essential to creating an account of the reference system and worldview of the Involved.   

In the case of Sidewalk Toronto, justification break-offs occur in the way that control over the system is to be exercised. Questions about the first kind of boundary judgment can be asked about not only the planning process but the design of the system itself: who controls different aspects of the system- will it be existing government authorities? Citizens? Sidewalk? Markets? Answering these questions in any manner prompts a second order boundary question: why should the smart city be administered in this way?

\subsubsection{Contrast complementary perspectives}
Critical reflection on the boundaries set out in an official storyline need not occur in a vacuum. We draw from the mainstream criticisms of Sidewalk in the press that speak to specific parts of the official story, and that often clearly reference the boundary judgments stabilized within the CSH framework. These complimentary perspectives offer avenues for critical reflection, suggest alternative ways of interpreting the situation, and provide support and direction for suspicions that arise from a perception of marginalization or coercion within the official boundaries.

Take for example the commitment to `complete communities'. Its key pillar is a mixed-income distribution in the Quayside, to be assured and measured through the provision of ``affordable housing''. The operationalization of affordability suggests how complete, mixed-income communities in Quayside are to be understood. At first glance, Sidewalk Lab's commitment to affordable housing seems to exceed the legal minimum set by municipal policy. Following a Sidewalk Fellows report, as well as criticism by community organization ACORN, they expanded the plan to include 40\% affordable housing\cite{fellows2018report,acorn2018acorn,waterfront2018table4}. 

However, this commitment does not address the affordable housing crisis in Toronto, it merely sets the floor based on the existing need\cite{rental2018Feb}. Crucially, WT and SL have operationalized affordability using the definition of the City of Toronto, which defines affordable housing as \emph{housing provided at a cost below the market rate}. Housing advocates and the Canadian Mortgage and Housing Corporation generally consider affordability as a relation between income and rent expenditure, i.e. less than 30\% \cite{cmhc2018about,pagliaro2018tornto}. More than one third of Toronto's residents cannot afford housing at current market rates. The market-based definition of affordable housing has changed the presumed beneficiaries of affordable housing in Sidewalk Toronto; now half of the affordable units are intended for low income residents, while the other is earmarked for ``middle income'' residents \cite{waterfront2018table4}. The remaining 60\% of units are for rental or sale at market rate, benefiting the affluent. This prioritization of the real-estate market at the expense of lower and middle income families and other renters demonstrates not only the worldview of involved parties, but the stakes of smart city development.

\subsubsection{Evaluate discourse and marginalization}
CSH can also allow reflection on where and how concerns about boundary judgments are marginalized. The specific boundaries drawn through the design, architecture, management, and governance of ICTs remain underdeveloped, but as CSH can focus equally well on process as on product, some of these boundaries are already tangible. The case of the Civic Data Trust (CDT) demonstrates how boundaries are drawn based on the worldview of the Involved, which focuses on innovation and resource allocation through markets, frustrating the honest consideration of smart city residents as beneficiaries. 

Much criticism of Sidewalk Toronto has focused on the tensions between developing smart cities to serve local publics versus the need to build them with a profit-oriented ``business model'' that allows SL to accrue value through the operation of the smart city. In part, discussion has focused on how the key resource of data generated in Quayside should be considered as a social product or public good, how the data is treated and made accessible, and to whom it belongs \cite{bean2017world,rattan2018torontonians,wylie2018sidewalk}. 

To address these concerns, decisionmakers announced a plan to create a ``Civic Data Trust'' (CDT), a ``third party public organization'' to govern a repository of Quayside Data \cite{waterfront2018table4,fellows2018report}. The CDT is intended to balance the goals of spurning innovation and protecting privacy, and to ``[safeguard] the public interest''. Itis meant to do so by committing to open standards in architecture and application programming interfaces (APIs), and by requiring data collection and sharing be minimal, done with `meaningful consent', and in accordance with existing laws. The governing body of the CDT balances the market value of this data with commitments to the people who are at once the source of this data, and its presumed beneficiaries. 

The proposal of the CDT was met with criticism. Former Privacy Commissioner for Ontario and global privacy expert Ann Cavoukian resigned from the aforementioned expert panel, citing unacceptable weaknesses in the fact that it did not require anonymization at the point of collection and merely \textit{encouraged} services collecting data to adhere to strong privacy principles \cite{cbc2018not}. While Cavoukian's resignation addressed concerns about privacy and social norms, it did not touch on the deeper issues of ownership and control. In the case of the CDT, this means elaborating on how governance decisions are made about authorization and licensing for data use, guiding principles, and compliance-- and by whom. These concerns are well spoken to by digital rights activist Bianca Wylie, who criticizes the way that value-laden concepts are deployed in the engagement process but do not lead to a more honest discussion about ownership and control over the direction and legal framework of constructs like the CDT\cite{wylie2018google}.

Decisions of ownership and control over technological infrastructure in the smart city draw boundaries with enormous social and economic impacts within these systems. The CDT would control the infrastructure of data storage and use, produced by the sensors that collect that data and the technical standards that structure access. The former head of Blackberry maker RIM summarized the problem of private ownership of the low level data infrastructure: because it enshrines Sidewalk's ownership of IP based on the collection of public data, it inevitably ``creates a systemic market advantage from which companies can inexorably expand'' \cite{ballsillie2018sidewalk}. Basillie argues that the ownership structure makes SL the major intended beneficiary of Quayside data, to say nothing of their relationship with Alphabet, whose major competencies and source of value are the exploitation of data. Wylie builds on concerns about this kind of ``platform capitalism'' in her critique of SL's ownership over the low level, built infrastructure, the data-layer infrastructure of storage and collection, as well as the access-layer that allows use of the data \cite{wylie2017civic}. Technological development making use of these layers could effectively be controlled by the strategic interest of Sidewalk Labs, potentially leading to the exclusion of local firms and community technologists from key development opportunities \cite{bean2017world}. 

SL has drawn their boundaries of control of resources very broadly. Tenuous agreements set out in the PDA protect the IP interests of Alphabet and Sidewalk, but do not provide much detail on how contracts for services might be decided in the future. Alphabet already controls applicable services such as \emph{Coord} (for managing transit and road-based assets) and has proposed numerous application level products for pilot at Quayside \cite{smyth2018announcing,doctoroff2017vision}. With control over the smart city architecture and their role as decision maker, SL appears positioned to place their own subsidiaries into market niches that they themselves would create, maintain, ad control. 

Control over the low-level technical and physical infrastructure by a single dominating firm allows a standards-level monopoly to shape the environment and market within which technology development and service innovation occurs. What role will those affected, the citizens and residents of Smart Cities, play in the decision making and governance of these new ways of life? Civic engagement and community building are alluded to in Sidewalk Toronto materials; Sidewalk says that its smart city will make volunteering easier, or that its data-layers will be a great resource for civic tech and social entrepreneurialism. However, these civic activities are situated within the boundaries of information systems controlled by Sidewalk. Examples include \emph{`Intersection'}, a product that provides internet access and ``enables a vast array of neighbourhood experiences, including amenity reservation and digital feedback channels''\cite[p.19]{doctoroff2017vision} and a ``Neighbourhood Assistant Tool'' that could ``enable Quayside residents to form new neighbourhood groups, crowd-source community needs, and access a peer-to-peer marketplace\dots another portal through which residents communicate feedback to officials, addressing the need for digital tools that gauge community well-being''  \cite[p.33]{doctoroff2017vision}.  

By juxtaposing official and critical discourses of the project, CSH allows us to see how opposing views are addressed, reconciled, and marginalized by the Involved. Specifically, we have tried to show how the \emph{Vision} for Sidewalk Toronto posits the company's role as a facilitator of civic engagement and democratic decision making, but not as the site or object of governance. Public engagement exercises seem to address governance issues, but in reality they obscure them. A CDT is just an act of trust if it does not have a governance structure that allows the civic body that it describes to exercise control over its data in decision making. While the proposed information systems seem to encourage civic engagement, their design also specifies what counts as valid engagement, suggesting that these concessions are a form of what Cardullo and Kitchin (via Arnstein) call ``placation'', where citizens are given the chance to change or challenge plans, but where their involvement is subordinated to the larger objectives of of those Involved\cite{cardullo2019being}.

\subsubsection{Identify boundary judgments}
The examples of affordable housing and the CDT illustrate some of the central motivations and boundary judgments that appear to be guiding the planning process. Tensions between resource collection and control are tied to boundary judgments around what constitutes the environment of smart city developments. For entities like WT and SL, the market is the ultimate environmental constraint. This is apparent, on one hand, by the way WT positions itself as a partner that seeks to leverage relationships with the private sector \cite{waterfront2018table3}. Their role as a steward of public lands working for complete communities is feasible only insofar as they attract capital investment, which is only possible if their partners can see a return on that investment \cite{waterfront2017rfp}. On the other hand, SL must look at smart city development as a way of maintaining competitive advantage, protecting its assets by maintaining control and authority in smart city governance. 

CSH has allowed us to stabilize an `official' story for the smart city planning process, and to leverage some existing, but marginalized, criticisms of that process to provide greater nuance to our perception of boundary judgments. We use the official narrative as a source of evidence for the way that these boundaries are constructed and maintained, and in some cases, how they serve to brush aside criticisms by giving the sense that they have been addressed, while still maintaining the same strategic boundaries. 

\subsubsection{Structure critique}
Table \ref{tab:CSH-SL} briefly summarizes key standpoints, concepts, and questions that arose during our analysis of official and critical perspectives. In the spirit of CSH, it should be seen as heuristic device rather than a result: not the outcome of CSH's application, but a mid-point that both helps to settle our reflections and points to more questions and avenues for critique. More iterations and versions of such tables can be produced individually or cooperatively. The table here represents a period of research and analysis, it serves as a summary of our thoughts at the time, as a provocation to further reflection and analysis within the categories of CSH. 

\begin{table*}\scriptsize
\setlength{\extrarowheight}{1pt}%
\label{tab:CSH-SL}
\caption{CSH Map Example}
\begin{tabular}{|p{3cm}|p{5cm}|p{3cm}|p{5cm}|}
\hline
\textbf{Category} & \textbf{Stated is} & \textbf{Selected  Concepts Relevant for Critique} & \textbf{Critical Standpoints} \\
\hline
1. Who is/ought to be the intended \textbf{beneficiary} of Sidewalk Toronto? & Create  ``a vibrant, climate-positive and prosperous {[}mixed-income{]} community\dots [a] model to encourage market transformation towards climate-positive city building''  by adopting a ``triple bottom line'' approach that balances environmental, economic, and ``socio-cultural'' purposes of people, profits, and planet \cite{waterfront2017rfp}. 
& \begin{itemize}[topsep=0pt, leftmargin=5pt, labelindent=0pt, itemindent=0pt]
    \item Complete communities 
    \item Transformative Sustainability 
    \item Sustainability and community within a market
\end{itemize} 
& \begin{itemize}[partopsep=0pt,leftmargin=5pt, labelindent=0pt, itemindent=0pt] 

\item Actual purpose observable in behavior \cite{meadows2008thinking}
\item Manifests in beneficiaries, actions, measures
\item Examine how control, knowledge, legitimation are handled to infer purpose
\item Deviations: consider e.g. affordable housing \end{itemize} \\
\hline

2. What is/ought to be the \textbf{purpose} of the Sidewalk Toronto? & Residents `of all income levels' get `complete communities' where they can live, work, and play in well-connected beautified space. \newline The project partner gets  an `unparalleled testbed environment' to `showcase advanced technologies' and  `financial opportunities' from development.\newline
Employers get  access to skilled labour, an emerging industry cluster, and local commercial opportunities.\newline
Visitors get access to public space and amenities.
& \begin{itemize}[topsep=0pt, leftmargin=5pt, labelindent=0pt, itemindent=0pt] 
\item Affordable housing defs
\item Privacy by Design
\item Value creation from smart city data
\item Viability of local firms \end{itemize} 
& \begin{itemize}[topsep=0pt, leftmargin=5pt, labelindent=0pt, itemindent=0pt]
\item Residents as data source for surveillance capitalism
\item `Affordable housing’ is (re)defined so that it becomes unaffordable for most
\item The interests of Sidewalk compete with those of other intended beneficiaries \end{itemize} \\
\hline

3. What is/ought to be the \textbf{measure of the improvement} of Sidewalk Toronto? 
& Under developement, described in terms of: 
\begin{itemize}[topsep=0pt, leftmargin=5pt, labelindent=0pt, itemindent=0pt]
\item Social goals met through thresholds for affordable housing and amenities
\item Environmental goals met through coherence with WT Resilience and Innovation Framework for Sustainability and LEED certification\cite{cgbc2018leed,waterfront2018resilience}
\item Economic goals met through investment returns -- economic output, government revenue, ``full-time employment years'', private sector investment
\end{itemize}
& \begin{itemize}[topsep=0pt, leftmargin=5pt, labelindent=0pt, itemindent=0pt]
\item Sustainability indicators\cite{bell2012sustainability}
\item Definitions of affordable housing
\item Tradeoffs  or conflicts between the 3Ps
\item Transactionalization of relations \end{itemize} 
& \begin{itemize}[topsep=0pt, leftmargin=5pt, labelindent=0pt, itemindent=0pt]
\item{Environmental sustainability as a service}
\item{affordable housing for the middle class} \end{itemize}\\
\hline

4. Who is/ought to be the \textbf{decision maker} in control of the resources for Sidewalk Toronto? & PDA establishes WT and SL \cite{waterfront2018plan}, under auspices of government & \begin{itemize}[topsep=0pt, leftmargin=5pt, labelindent=0pt, itemindent=0pt]
\item Boundaries around governance
\item `Public Interest' as a contested concept \end{itemize}
& \begin{itemize}[topsep=0pt, leftmargin=5pt, labelindent=0pt, itemindent=0pt]
\item CDT governance left outside boundaries, proposed information systems enclose civic engagement in decision making.
\item Role of public engagement gives WT, SL the initiative in communication and final calls.\end{itemize} \\
\hline

5. What \textbf{resources} are/ought to be under the control of Sidewalk Toronto? 
& \begin{itemize}[topsep=0pt, leftmargin=5pt, labelindent=0pt, itemindent=0pt]
\item Data: Civic Data trust (CDT)
\item IP: set out in MIDP
\item Property: largely left up to the market
\end{itemize}
& \begin{itemize}[topsep=0pt, leftmargin=5pt, labelindent=0pt, itemindent=0pt]
\item CDT governance
\item Platform Capitalism \cite{srnicek2017platform}\end{itemize} 
& Criticism by Ann Cavoukian around IP points to discretionary power of SL around publicly produced data\cite{cbc2018not}, and ownership of the standards layer positions SL for bottom level control \cite{ballsillie2018sidewalk} \\
\hline

6. What is/ought to be outside the control of the decision maker, the \textbf{environment}? 
& \begin{itemize}[topsep=0pt, leftmargin=5pt, labelindent=0pt, itemindent=0pt]
\item Markets, investor capital, competitive advantage
\item Public opinion \end{itemize} 
& Public/Private Partnerships and the enclosure of civic engagement
& Community is not based on consumption of the same services but based on interdependence and complex relationships among members \\
\hline

7. Who is/ought to be the \textbf{experts} providing the relevant knowledge and skills ? 
& \begin{itemize}[topsep=0pt, leftmargin=5pt, labelindent=0pt, itemindent=0pt]
\item Public Participation
\item SL and WT experts \end{itemize} 
& \begin{itemize}[topsep=0pt, leftmargin=5pt, labelindent=0pt, itemindent=0pt]
\item{``Tokenistic'' engagement \cite{cardullo2019being}}
\item{Feedback as user testing}
\item{Utopian aesthetics in \emph{Vision} and promotional material\cite{sauter_googles_2018}}
\end{itemize}
& Technical/professional administration vs situated knowledge \\
\hline

8. What are the relevant knowledges and skills (\textbf{expertise}) necessary for the operation of Sidewalk Toronto? 
& Data science and ICT: ``the only urban innovation company built expressly to bridge the divide \dots between urbanists and technologists\dots No one
else has envisioned the integration of technology into the physical
environment that will give rise to an urban innovation platform\dots''\cite[p.16]{doctoroff2017vision} 
& \begin{itemize}[topsep=0pt, leftmargin=5pt, labelindent=0pt, itemindent=0pt]
\item Publics\cite{disalvo2009design,dewey2012public} as the object of data collection
\item Everyday expertise, situated knowledges \end{itemize} 
& \begin{itemize}[topsep=0pt, leftmargin=5pt, labelindent=0pt, itemindent=0pt]
\item ``Confusion over what codesign means''\cite{wylie2018sidewalk2}. 
\item Public  engagements  seen not as  a  source  of expertise, but as SL leveraging public engagements as a source of knowledge for their own experts, to `inform' MIDP \cite{sidewalk2018djbackground}.\end{itemize} \\
\hline

9. What \textbf{guarantees} the successful implementation of Sidewalk Toronto? 
& \begin{itemize}[topsep=0pt, leftmargin=5pt, labelindent=0pt, itemindent=0pt]
\item Government, WT as steward
\item Sidewalk Labs~/Alphabet 
\end{itemize}
& \begin{itemize}[topsep=0pt, leftmargin=5pt, labelindent=0pt, itemindent=0pt]
\item Technological Solutionism
\item Faith in markets
\end{itemize}
& \begin{itemize}[topsep=0pt, leftmargin=5pt, labelindent=0pt, itemindent=0pt]
\item Democracy as Guarantor
\end{itemize} \\
\hline

10. Who is/ought to be considered a \textbf{witness} representing the interests of those affected by, but not involved with Sidewalk Toronto? 
& \begin{itemize}[topsep=0pt, leftmargin=5pt, labelindent=0pt, itemindent=0pt]
\item Public Engagement Process, including panels, Roundtables, and workshops
\item Appeals to Indigenous Planning in resident’s panel and Roundtables \cite{sidewalk2018rrp,waterfront2018table4}\end{itemize} 
& \begin{itemize}[topsep=0pt, leftmargin=5pt, labelindent=0pt, itemindent=0pt]
\item Curated publics \end{itemize}
& Atomization and aggregation poses public as individuals to provide affective feedback and to components rather than the system. Focus on values provides rhetorical initiative to WT and SL. No apparent inclusion or reference beyond land acknowledgement \\
\hline

11. What are/ought to be the opportunities for the interests of those affected free themselves from (\textbf{emancipation}) the worldview of Sidewalk Toronto? 
& Holistic Planning,  where ``innovation, community priorities, policy objectives, placemaking, phasing, infrastructure, economics, market, site planning, and technical issues will be thoughtfully merged.'' \cite[p.59]{doctoroff2017vision} 
& \begin{itemize}[topsep=0pt, leftmargin=5pt, labelindent=0pt, itemindent=0pt]
\item Highly structured process
\item Feedback as representation/engagement
\end{itemize}
& Preconfigured engagement, based on identified areas, and offering no potential for emancipation \\
\hline 

12. What space is/ought available for reconciling different \textbf{worldviews} regarding Sidewalk Toronto among those affected but not involved? 
& Private entities, market transformation, in the ``most measurable community in the world''\cite{doctoroff2017vision}  
& \begin{itemize}[topsep=0pt, leftmargin=5pt, labelindent=0pt, itemindent=0pt]
\item Politics as management
\item Computational thinking \cite{easterbrook2014computational}
\item City as computer \cite{mattern2017city}\end{itemize} 
& Appealing to the `reality' of neoliberal dominance, the balance of the 3Ps tilting to profit, which is always the most bottom line.\\
\hline
\end{tabular}
\end{table*}
\section{Discussion}
\subsection{Reflections and Limitations}
As our tabular application of CSH (\ref{tab:CSH-SL}) demonstrates, the authors' standpoint and the requirements of the format have acted to select and constrain the perspective of this partial engagement with the Sidewalk Toronto planning process. Our account here is not comprehensively holistic in terms of description. The reproduction of urban life relies on global and local systems well beyond our scope here. As a document of process, the CSH table passes over many considerations and commitments to matters of environmental, social, economic, and individual justice.

In application, CST and CSH research could anticipate these limitations and address them with a systematic expansion of methods. We have collected mainstream press and official documentation, and attended events when possible. It would be desirable to \emph{sweep in} a more diverse array of collaborators, participants, and publics into the creation of CSH documentation\cite{churchman1968west}. In matters of public concern, CSH is always one voice among many. As Systemic Intervention \cite{midgley_systemic_2000} or other action research, applying CSH could help maintain space for critical reflection and diversity of perspective.

\subsection{Measured Management}
``Sidewalk expects Quayside to become the most measurable community in the world'' \cite[pg.22]{doctoroff2017vision}. Ambitions to environmental sustainability in Sidewalk Toronto rest on the belief that through measurement and systematization of urban activities such as energy use and transit, control and coordination can be improved to lessen waste and environmental impact. Smart City technology enables pursuit of this goal through sensory and management systems that might track and analyze trends, more efficiently consuming resources. In the case of a smart city like Sidewalk, these resources might include stocks for heating, waste and recycling capacity, and for maintaining the physical spaces sought after for conducting economic and social activity, including dwelling, commercial, and public spaces like parks. If discrete or aggregated measurements can be sought, monitored, and managed, that data stands in as a resource. As a representational layer on top of the city's (not to mention its associates) material activities, smart city systems want to manage that data. Value creation promised through increased convenience, efficiency, sustainability, etc., are delivered only if the presumed requisite data is available, and on who has access to it. Even after construction, capital resources and labour are required needed to maintain these infrastructures and to put their results to use.

Ambitions in the \emph{Vision} extend beyond environmental sustainability, to ``promoting activity, healthy eating, relaxation, and connection to the environment\dots [through] capturing a variety of data and facilitating residents' and others' uses of that data through existing and new applications'', as well as to the business of neighbourhood politics \cite[pg.171]{doctoroff2017vision}. From this perspective, ecological sustainability intersects with other aspects of sustainability as resource consumption and use is related with activities, lifestyles, exchanges (etc.), and with supporting infrastructures, computational and otherwise. However, mere measurement is insufficient for management: ``to understand what makes the urban environment work well, and detect when it is under-performing, it is necessary to perform longitudinal analysis, and be able to distinguish normal states from anomalous ones'' \cite[pg.72]{doctoroff2017vision}.

Achieving normal, ``livable'' space requires evaluation, and a value-based framework that is implied, but, we think, unspecific. Sidewalk's commitment to Quayside suggests an ongoing endeavour, actively managed at different time scales with successive interventions by, for example, ``facility managers'' \cite[pg.74]{doctoroff2017vision}. Commercially, efficiency and convenience can perform as MoIs-- a decrease in travel time or noise complaints, increases in overall air quality, successful user transactions, etc. Proposed innovation platforms or information marketplaces articulate the resident/citizens of a smart city primarily as the beneficiaries of services, or consumers of resources. Just who delivers these services and manages the resources? Would municipal or public actors be replaced by a market where returns were sought through treating residents and citizens as a resource or product \cite{cardullo2019being}? Politically, objects like the CDT or smart city MoIs could be a site for citizen control. A democratic approach to smart city design would make citizen involvement in the creation of these standards a part of the planning process, open for debate and driven by the values of residents. Already within the context of smart city technologies, Balestrini et al., have used the concept of a ``city commons'' to guide participatory design of a sensor network\cite{balestrini2017city}. So far, however, Sidewalk Toronto's planning process represents publics without including them. 
\section{Conclusions}
\label{sec:org963d911}

\begin{quoting}[font=itshape,vskip=.5em]
\noindent`Cities have the capability of providing something for everybody, only because, and only when, they are created by everybody.'~\cite{jacobs_death_1961}
\end{quoting}

The ongoing SL project in Toronto is a major smart city development, the deployment of technologically advanced urban infrastructures built from scratch to achieve their designers' ambitious goals. Sustainability is a key value used to motivate the project. As this paper argued, however, sustainability in smart city projects must be considered more holistically than is possible through the narrow lenses of technological optimization and ecological sustainability\cite{becker2015sustainability}. 

Basing our analysis on CST, we have showed how the project's purpose is framed by the rationalities and goals of its most powerful stakeholders. We used CSH for a boundary critique that traces the concerns of the numerous voices, and to examine and critically reflect on how values and interests influence and constrain the project's purposes and vision. This situates CSH, and CST more broadly, as a powerful tool for the holistic consideration of sustainability in smart city projects and large scale transitions to sustainable societies.

This is not a call to disregard the technical expertise offered by Sidewalk or any other firm. If we are to truly transform our societies to be sustainable and prosperous, we must critically reflect on the role and purpose of technologies in our everyday lives. What becomes clear through critical analysis is that technical decisions in the smart city are, as much as ever, political decisions about relations between people, organizations, and power. Governance of smart cities is tied to design and architecture choices, both in regards to the sustainability of any underlying infrastructure of smart cities, but also to the intersection of this new datafied layer and everyday life, through the systems and interfaces constructed to create and manage that resource.

A key challenge for the ICT4s community is to ensure that these types of boundaries do not go unnoticed in smart city research. This will mean continuing to critically reflect on the role of technological choices in instantiating social and political relationships and assessing their potential as environmentally sustainable technologies. Critical systems thinking can assist in just that type of work. As a heuristic tool, CSH does not need to supplant or dominate research in order to be effective. Beyond this evaluative role, as framework for interventionary practise or self-reflection in technology design it can help practitioners and technologists to more effectively probe the normative implications of their work. 

\section*{Acknowledgements}
This research was partially supported by Natural Sciences and Engineering Research Council through RGPIN-2016-06640, the Canada Foundation for Innovation, Onatrio Research Fund, and the Social Science and Humanities Research Council through the  Canadian Graduate Scholarship. Special thanks to Dawn Walker.

\bibliographystyle{ieeetr}
\bibliography{sl-csh}

\begin{thebibliography}{10}

\bibitem{doctoroff2017vision}
D.~L. Doctoroff, ``Sidewalk toronto project proposal,'' proposal, Sidewalk
  Labs, 10 2017.

\bibitem{sauter_googles_2018}
M.~Sauter, ``Google’s {Guinea}-{Pig} {City},'' {\em The Atlantic}, Feb. 2018.

\bibitem{editorial_guardian_2018}
{The Guardian}, ``The {Guardian} view on {Google} and {Toronto}: Smart city,
  dumb deal,'' Feb. 2018.
\newblock Editorial.

\bibitem{mann2018shifting}
S.~Mann, O.~Bates, and R.~Maher, ``Shifting the maturity needle of ict for
  sustainability,'' {\em Proc. 5th Int Conf on ICT4S}, 2018.

\bibitem{easterbrook2014computational}
S.~Easterbrook, ``From computational thinking to systems thinking,'' in {\em
  Proc. 2nd Int Conf on ICT4S}, 2014.

\bibitem{becker2015sustainability}
C.~Becker, R.~Chitchyan, L.~Duboc, S.~Easterbrook, B.~Penzenstadler, N.~Seyff,
  and C.~C. Venters, ``Sustainability design and software: The karlskrona
  manifesto,'' in {\em Proc., 37th Intl. Conference on Software Engineering},
  vol.~2, IEEE Press, 2015.

\bibitem{kamilaris2015social}
A.~Kamilaris, A.~Pitsillides, C.~Fidas, and S.~Kondepudi, ``Social electricity:
  The evolution of a large-scale, green ict social application through two case
  studies in cyprus and singapore,'' in {\em Proc. 3rd Int Conf of ICT4S},
  (Copenhagen, Denmark), Atlantis Press, 2015.

\bibitem{von2015citizen}
P.~F. von Heland, M.~Nyberg, A.~Bondesson, and P.~Westerberg, ``The citizen
  field engineer: Crowdsourced maintenance of connected water infrastructure,''
  in {\em Proc. 3rd Int Conf on ICT4S}, Atlantis Press, 2015.

\bibitem{shahrokni2014big}
H.~Shahrokni, B.~Van~der Heijde, D.~Lazarevic, and N.~Brandt, ``Big data gis
  analytics towards efficient waste management in stockholm.,'' in {\em Proc.
  2nd Int Conf on ICT4S}, Atlantis Press, 2014.

\bibitem{brylka2014ict}
J.~Studzinski, R.~Brylka, and K.~Kazubski, ``Ict system for smart city
  management,'' in {\em Proc. 2nd. Int Conf on ICT4S}, Atlantis Press, 2014.

\bibitem{kramers2018aaas}
A.~Kramers, T.~Ringenson, L.~Sopjani, and P.~Arnfalk, ``Aaas and maas for
  reduced environmental and ccclimate impact of transport,'' in {\em Proc. 5th
  Int Conf on ICT4S}, vol.~52, EasyChair, 2018.

\bibitem{cosgrave2014smart}
E.~Cosgrave, T.~Tryfonas, and T.~Crick, ``The smart city from a public value
  perspective,'' in {\em Proc. 2nd. Int Conf on ICT4S}, Atlantis Press, 2014.

\bibitem{kramers2014planning}
A.~Kramers, J.~Wangel, and M.~H{\"o}jer, ``Planning for smart sustainable
  cities,'' in {\em Proc. 2nd Int Conf on ICT4S}, Atlantis Press, 2014.

\bibitem{borjesson2015ict}
M.~B{\"o}rjesson~Rivera, E.~Eriksson, and J.~Wangel, ``Ict practices in smart
  sustainable cities: In the intersection of technological solutions and
  practices of everyday life,'' in {\em Proc. EnviroInfo and ICT for
  Sustainability 2015}, (Copenhagen), Atlantis Press, 2015.

\bibitem{waterfront2017rfp}
{Waterfront Toronto}, ``Request for proposals innovation and funding partner
  for the quayside development opportunity,'' Request for Proposal 2017-13,
  Waterfront Toronto, 03 2017.

\bibitem{jackson2003systems}
M.~C. Jackson, {\em Systems thinking: Creative holism for managers}.
\newblock Citeseer, 2003.

\bibitem{ulrich1983critical}
W.~Ulrich, {\em Critical Heuristics of Social Planning: A New Approach to
  Practical Philosophy}.
\newblock J. Wiley and Sons, 1983.

\bibitem{ulrich2010critical}
W.~Ulrich and M.~Reynolds, ``Critical systems heuristics,'' in {\em Systems
  Approaches to Managing Change: A Practical Guide}, Springer, 2010.

\bibitem{reynolds2007evaluation}
M.~Reynolds, ``Evaluation based on critical systems heuristics,'' 2007.

\bibitem{henriques_triple_2004}
A.~Henriques and J.~Richardson, eds., {\em The Triple Bottom Line: Does it all
  Add Up? : Assessing the Sustainability of Business and {CSR}}.
\newblock Earthscan Publications Ltd., 2004.

\bibitem{waterfront2018plan}
{Waterfront Toronto}, ``Plan development agreement between toronto waterfront
  revitalization corporation and sidewalk labs llc,'' agreement, {Waterfront
  Toronto}, 07 2018.

\bibitem{residents2018report}
F.~Chowdhury, F.~West, G.~Chau, G.~Hewitt, G.~Cabigas, H.~Kang, H.~Mactaggart,
  J.~Goodman, J.~Duong, J.~Ruzic, J.~Slaughter, K.~Donaldson, L.~Clamageran,
  M.~Houghron, P.~MacKay, R.~Jia, S.~Mahmood, S.~Mehta, and S.~Fung, {\em
  Resident's Reference Panel Interim Report}.
\newblock Sidewalk Toronto, 2018.

\bibitem{fellows2018report}
A.~Wishloff, A.~Espanol, B.~Chang, C.~Leung, C.~Yeung, H.~Brath, H.~Ngo, K.~S.
  Louis-McBurnie, P.~Seufert, S.~Persaud, S.~Chan, and W.~Sutter, {\em Sidewalk
  Toronto Fellow Report}.
\newblock Sidewalk Toronto, 2018.

\bibitem{kids2018report}
{Sidewalk Toronto} and {Kids Camp Participants}, {\em Sidewalk Toronto Summer
  Kids Camp}.
\newblock Sidewalk Toronto, 2018.

\bibitem{sidewalk2018youtube}
{Sidewalk Toronto}, 2018.
\newblock YouTube Channel.

\bibitem{ulrich1987critical}
W.~Ulrich, ``Critical heuristics of social systems design,'' 1987.

\bibitem{acorn2018acorn}
{ACORN Canada}, ``Acorn to sidewalk toronto: Give us real affordable
  housing!.'' Online, 2018.

\bibitem{waterfront2018table4}
{Waterfront Toronto and Sidewalk Labs}, ``{Public Roundtable \#4}.''
  Presentation, 12 2018.

\bibitem{rental2018Feb}
{City News Staff}, ``{By the numbers: Toronto's rental market - CityNews
  Toronto},'' Feb 2018.
\newblock [Online; accessed 21. Jun. 2018].

\bibitem{cmhc2018about}
{Canadian Mortgage and Housing Corporation}, ``About affordable housing.''
  Online, 04 2018.

\bibitem{pagliaro2018tornto}
J.~Pagliaro, ``Toronto to review defintion of `affordable' housing,'' {\em The
  Toronto Star}, 2018.

\bibitem{bean2017world}
B.~Bean, ``The world is watching as data drives toronto’s smart city
  experiment,'' {\em Open Data Exchange}, 10 2017.

\bibitem{rattan2018torontonians}
C.~Rattan, ``Torontonians should take control of their data,'' {\em Now
  Magazine}, 05 2018.

\bibitem{wylie2018sidewalk}
B.~Wylie, ``Sidewalk toronto, procurement innovation, and permission to fail,''
  {\em Medium}, 04 2018.

\bibitem{cbc2018not}
{CBC News}, ```not good enough': Toronto privacy expert resigns from sidewalk
  labs over data concerns,'' {\em CBC News}, 10 2018.

\bibitem{wylie2018google}
B.~Wylie, ``Google is still planning a `smart city' in toronto despite major
  privacy concerns,'' {\em Motherboard}, August 2018.

\bibitem{ballsillie2018sidewalk}
J.~Balsillie, ``Sidewalk toronto has only one beneficiary, and it is not
  toronto,'' {\em The Globe and Mail}, 10 2018.
\newblock Editorial.

\bibitem{wylie2017civic}
B.~Wylie, ``{Civic Tech: On Google, Sidewalk Labs, and Smart Cities},'' {\em
  Torontoist}, Oct 2017.

\bibitem{smyth2018announcing}
S.~Smyth, ``Announcing coord: The integration platform for mobility providers,
  navigation tools, and urban infrastructure,'' {\em Sidewalk Labs}, 2018.

\bibitem{cardullo2019being}
P.~Cardullo and R.~Kitchin, ``Being a ‘citizen’in the smart city: up and
  down the scaffold of smart citizen participation in dublin, ireland,'' {\em
  GeoJournal}, vol.~84, no.~1, pp.~1--13, 2019.

\bibitem{waterfront2018table3}
{Waterfront Toronto and Sidewalk Labs}, ``{Public Roundtable \#3}.''
  Presentation, 08 2018.

\bibitem{meadows2008thinking}
D.~H. Meadows and D.~Wright, {\em Thinking in Systems: A Primer}.
\newblock Chelsea Green publishing, 2008.

\bibitem{cgbc2018leed}
{Canada Green Building Council}, ``Leed\textregistered certification process.''
  Online, 2018.

\bibitem{waterfront2018resilience}
{Waterfront Toronto} and {Canadian Urban Institute}, ``Waterfront toronto
  resilience and innovation framework for sustainability,'' tech. rep.,
  Waterfront Toronto, 2017.

\bibitem{bell2012sustainability}
S.~Bell and S.~Morse, {\em Sustainability Indicators: Measuring the
  Immeasurable?}
\newblock Routledge, 2012.

\bibitem{srnicek2017platform}
N.~Srnicek, {\em Platform capitalism}.
\newblock John Wiley \& Sons, 2017.

\bibitem{disalvo2009design}
C.~DiSalvo, ``Design and the construction of publics,'' {\em Design Issues},
  vol.~25, no.~1, 2009.

\bibitem{dewey2012public}
J.~Dewey, {\em The Public and its Problems: An Essay in Political Inquiry}.
\newblock Penn State Press, 2012.

\bibitem{wylie2018sidewalk2}
B.~Wylie, ``Sidewalk toronto, social license, and the limits of a borrowed
  reputation,'' {\em Medium}, 06 2018.

\bibitem{sidewalk2018djbackground}
{Sidewalk Toronto}, ``{Design Jam: Background}.'' Presentation, 2018.

\bibitem{sidewalk2018rrp}
{Sidewalk Toronto}, {\em Sidewalk Toronto Residents Reference Panel FAQ}.
\newblock Sidewalk Toronto, 2018.

\bibitem{mattern2017city}
S.~Mattern, ``A city is not a computer,'' {\em Places Journal}, 2017.

\bibitem{churchman1968west}
C.~W. Churchman, {\em The Systems Approach}.
\newblock Delta, New York, 1968.

\bibitem{midgley_systemic_2000}
G.~Midgley, {\em Systemic {Intervention}}.
\newblock Springer, 2000.

\bibitem{balestrini2017city}
M.~Balestrini, Y.~Rogers, C.~Hassan, J.~Creus, M.~King, and P.~Marshall, ``A
  city in common: a framework to orchestrate large-scale citizen engagement
  around urban issues,'' in {\em Proc. `17 CHI Conference on Human Factors in
  Computing Systems}, pp.~2282--2294, ACM, 2017.

\bibitem{jacobs_death_1961}
J.~Jacobs, {\em The {Death} and {Life} of {Great} {American} {Cities}}.
\newblock Vintage Books, 1961.

\end{thebibliography}
\end{document}